\documentclass[a4paper,11pt]{article}

\usepackage{amsmath}
\usepackage{rotating}
\usepackage{xypic}
\usepackage{a4wide}
\usepackage{amsfonts}
\newenvironment{pf}{\textbf{Proof:}}{\hspace{\stretch{1}}$\square$}

\newtheorem{thm}{Theorem}

\newtheorem{pr}{Proposition}

\font\timesept=cmr7
\title{Diffusion Approximation of Stochastic Master Equations with Jumps}
\author{ C. Pellegrini$^\dag$ and F. Petruccione$^\ddag$}

\begin{document}

\maketitle
\centerline{\timesept Scoool of Physics, University of KwaZulu-Natal}
\vskip -1mm
\centerline{\timesept and National Institute for Theoretical Physics}
\vskip -1mm
\centerline{\timesept Private Bag X54001}
\vskip -1mm
\centerline{\timesept Durban 4000, South Africa}
\vskip -1mm
\centerline{\timesept $\dag$ pelleg@math.univ-lyon1.fr}
\vskip -1mm
\centerline{\timesept $\ddag$ petruccione@ukzn.ac.za}
\begin{abstract}
In the presence of quantum measurements with direct photon 
detection the evolution of open quantum systems is usually described by stochastic master equations with jumps. Heuristically, diffusion models can be obtained from these equations as  approximation. A condition for a general diffusion approximation for jump master equations is presented. This approximation is rigorously proved by using techniques for Markov processes which are based upon the convergence of Markov generators and martingale problems. This result is illustrated by rigorously obtaining the diffusion approximation for homodyne and heterodyne detection.  
\end{abstract}

\section*{Introduction}

In open quantum systems \cite{c,francesco2,carm1,carm2,GardinerZoller,OQS}, especially in the theory of measurement, an active line of research concentrates on the study of \textit{stochastic master equations} and \textit{stochastic Schr\"odinger equations} \cite{BAR,Book,francesco2,MR1121822,MR1919502,MR1639788,barchielli0,belav2,belav3,belav4,SSEMB,Diosi1,Diosi2,p1,p2,p3,infinite,W1,W22,W2}. These equations are stochastic differential equations (SDEs) describing the evolution of a small system undergoing continuous measurement. Specifically, stochastic Schr\"odinger equation describe the evolution of vectors in Hilbert space (\textit{wave function} of the small system) \cite{Book,SSEMB,francesco2,barchielli0,infinite}. Moreover, the notion of stochastic 
Schr\"odinger equation is closely related to the concept of \textit{unravelling} in quantum mechanics which is highly used to derive \textit{Monte Carlo wave function simulation methods} (see Chapter 7 in \cite{Book} and Chapters 17-19 in \cite{carm2} for examples, applications and for other references). The term stochastic master equation refers to the evolution of density matrices (\textit{states}\footnote{A state or a density matrix is a positive trace class operator of trace one.} of the system) \cite{Book,MR1121822,SSEMB,p1,p2,p3,francesco2} (see \cite{Book,MR1639788,p1,p2} for the equivalence between these two types of equations). 

The framework sketched above has a wide range of applications in quantum optics and quantum communications \cite{BAR,Book,MR1639788,belav2,belav3,carm1,carm2,GardinerZoller,SSEMB,Diosi1,Diosi2,francesco2,MR2271425,moller,OQS,milb,W1,W2}. A particularly important application is the modeling of photon detection. A typical situation is the one of a small system, e.g., a two level atom, stimulated by a laser. The measurement scheme is the continuous detection of the fluorescence photons emitted by the two level system. In this situation, the evolution of the state of the system is given by the following stochastic master equation
 \begin{equation}\label{j1}
d\rho_t=L(\rho_t)dt+\left(\frac{C\rho_t C^\star}{\textrm{	\textrm{\textrm{Tr}}}\big[C\rho_t C^\star\big]}-\rho_t\right)\big(d\tilde{N}_t-	\textrm{\textrm{Tr}}\big[C\rho_t C^\star\big]dt\big),
\end{equation}
where $(\tilde{N}_t)$ is a counting process with stochastic intensity $\int_0^t	\textrm{\textrm{Tr}}\big[C\rho_s C^\star\big]ds.$ In this equation the operator $L$ is called Lindblad operator and takes the form
\begin{equation}
L(\rho)=-i[H,\rho]-\frac{1}{2}\{C^\star C,\rho\}+C\rho C^\star,
\end{equation}
where $C$ is an arbitrary operator. The Lindblad operator is actually the generator of the reduced dynamic of the qubit if there were no measurements. The occurrence of a jump in equation $(\ref{j1})$ corresponds to the emission of a photon by the qubit. Another type of equation describes diffusive evolutions, that is,
\begin{equation}\label{dif1}
d\rho_t=L(\rho_t)dt+\big(C\rho_t+\rho_t C^\star-\textrm{\textrm{Tr}}\big[\rho_t(C+C^\star)\big]\big)dW_t,
\end{equation}
where $(W_t)$ is a standard Brownian motion. Such an equation is related to heterodyne or homodyne detection (the first one allows to study the spectrum of the light and the second one the squeezing \cite{BAR,Book,francesco2,carm1,carm2,W1,W2}). 

Heuristically, equation (\ref{dif1}) can be obtained from the model (\ref{j1}) by the following approach \cite{BAR,SSEMB,francesco2,Gisin1,Gisin2,W1,W2}. In equation (\ref{j1}), we replace the operator $C$ by $C_\varepsilon=C+I/\varepsilon,$ and we define the process $(W_t^\varepsilon)$ by $dW_t^\varepsilon=\varepsilon d\tilde{N}_t-dt/\varepsilon$. In this way, formally by using usual \textit{Ito stochastic calculus} we get $dW_t^\varepsilon dW_t^\varepsilon=\varepsilon^2d\tilde{N}_t=\varepsilon dW_t^\varepsilon+dt$ and by taking the limit $\varepsilon$ goes to zero, we get $\lim_{\varepsilon\rightarrow0}W_t^\varepsilon=W_t$, where $W_t$ is a Brownian motion. Next, by assuming $C=C^\star$ and by taking the formal limit, $\varepsilon$ goes to zero, in
 \begin{equation}
d\rho_t=L_\varepsilon(\rho_t)dt+\frac{1}{\varepsilon}\left(\frac{C_\varepsilon\rho_t C_\varepsilon^\star}{\textrm{	\textrm{\textrm{Tr}}}\big[C_\varepsilon\rho_t C_\varepsilon^\star\big]}-\rho_t\right)\varepsilon\big(d\tilde{N}_t-	\textrm{\textrm{Tr}}\big[C_\varepsilon\rho_t C_\varepsilon^\star\big]dt\big),
\end{equation}
we recover the equation (\ref{dif1}). Such a result is called \textit{diffusion approximation} (see \cite{Champ, Const, Lee, Nag} for similar result in biology, classical mechanic or mathematical finance). Let us stress that the equations $(\ref{j1})$ and $(\ref{dif1})$ are particular cases of stochastic master equations. More general situations are described by jump-diffusion SDEs \cite{barchielli0,infinite,p3}. 

The main aim of this article is to prove rigorously the diffusion approximation. We address this question in the more general models of jump-diffusion SDEs. We present a general condition for the coefficients defining the SDEs which ensures diffusion approximation results. We show that this condition appears naturally in concrete physical applications.
\bigskip

The article is organized as follows. In Section $1$, we investigate the diffusion approximation problem for the jump-diffusion stochastic master equations. The approach is based on convergence of Markov generators and problems of martingale. Then, we obtain a  sufficient condition where the diffusion approximation is valid. In Section $2$, we show in concrete examples, that such situations appear naturally. In particular, we justify rigorously the diffusive models of homodyne and heterodyne detection in quantum optics.

\section{Diffusion Approximation of Jump-Diffusion Stochastic Master Equations}

\subsection{Framework and Main Result}

We consider a Hilbert space $\mathcal{H}=\mathbb{C}^N$, which represents the small system, which undergoes indirect quantum measurement. In order to describe the random evolution of the small system, we consider a probability space $(\Omega,\mathcal{F},\mathcal{F}_t,P)$, which supports a $n$ dimensional Brownian motion $(W_t=W_1(t),\ldots,W_n(t))$ and $k$ independent Poisson point processes $N_i(), i=1,\ldots,k$ on $\mathbb{R}^2$, independent from the Brownian motion. Let us remind that, for all Borel sets $A$ of $\mathbb{R}^2$, the quantity $N_i(A)$ corresponds to the number of points of the Poisson point process $N_i$ into the set $A$. Moreover, for each $i$, the application $A\longmapsto N_i(A)$ defined on the Borel sets defines a measure. More precisely, this defines a \textit{Random measure} \cite{CSPM}. Moreover, for all Borel sets $A$, we have $\mathbb{E}[N_i(A)]=\lambda(A)$, where $\lambda$ corresponds to the Lebesgue measure on $\mathbb{R}^2$ (here, the Lebesgue measure is called \textit{intensity measure} of the Poisson point processes \cite{CSPM}). Now, by denoting the differential elements as $N_i(dx,ds)$, we can consider the following stochastic differential equation
\begin{eqnarray}\label{generic}
&&\rho_t=\int_0^tL(\rho_{s\mbox{\tiny{-}}})ds+\sum_{i=1}^n\int_{0}^t \Big(C_i\rho_{s\mbox{\tiny{-}}}+\rho_{s\mbox{\tiny{-}}} C^\star-\textrm{\textrm{Tr}}\big[\rho_{s\mbox{\tiny{-}}}(C+C^\star)\big]\Big)dW_i(s)\nonumber\\
	&&+\sum_{i=1}^{k}\int_0^t\int_{\mathbb{R}}\left(\frac{D_i\rho_{s\mbox{\tiny{-}}}D_i^\star}{\textrm{\textrm{Tr}}\big[D_i\rho_{s\mbox{\tiny{-}}}D_i^\star\big]}-\rho_{s\mbox{\tiny{-}}}\right)\mathbf{1}_{0<x<\textrm{\textrm{Tr}}[D_i\rho_{s\mbox{\tiny{-}}}D_i^\star]}\Big(N_i(dx,ds)-\textrm{\textrm{Tr}}\big[D_i\rho_{s\mbox{\tiny{-}}}D_i^\star\big]dxds\Big),\nonumber\\
\end{eqnarray}
where $L$ is a Lindblad operator, defined by
\begin{eqnarray}\label{LL}
L(\rho)=-i[H,\rho]-\frac{1}{2}\sum_{i=1}^n\{C_i^\star C_i,\rho\}-\frac{1}{2}\sum_{i=1}^k\{D_i^\star D_i,\rho\}+\sum_{i=1}^nC_i\rho C_i^\star+\sum_{i=1}^kD_i\rho D_i^\star.
\end{eqnarray}
In the equation (\ref{LL}), the operators $C_i$, $i=1,\ldots,n$ and $D_i$, $i=1,\ldots,k$ are $N\times N$ matrices. The operator $H$ is a self adjoint operator called free Hamiltonian of $\mathcal{H}$. The equation (\ref{generic}) is a generic stochastic master equation (see \cite{barchielli0,infinite} for different versions). Let us stress that, by using a Poisson point process on $\mathbb{R}^2$, we can define rigorously the counting processes: 
$$\tilde{N}_i(t)=\int_0^t\int_\mathbb{R}\mathbf{1}_{0<x<\textrm{\textrm{Tr}}[D_i\rho_{s\mbox{\tiny{-}}}D_i^\star]}N_i(dx,ds),\,\,i=1,\ldots,k,$$ which are clearly counting processes with stochastic intensities $\int_0^t\textrm{Tr}[D_i\rho_{s\mbox{\tiny{-}}}D_i^\star]ds$. The solution of (\ref{generic}) is called a \textit{quantum trajectory} and takes values in the set of states of $\mathcal{H}$; this describes the random evolution of $\mathcal{H}$ in presence of continuous indirect measurement (see Refs. \cite{barchielli0,infinite,p1,p2,p3} for justifications on existence and uniqueness of solutions). In the rest of the paper we denote $\mathcal{S}$ the set of states, this is a compact subset of $\mathcal{B}(\mathcal{H})$.

Our aim consists in generalizing the heuristic rule described in the Introduction and to prove rigorously diffusion approximations in such models. The mathematical framework is the one of the convergence in distribution for stochastic processes. To this end, let $T>0$ be fixed, we introduce the space $\mathcal{D}[0,T[$, which corresponds to the set of c\`adl\`ag processes endowed with the Skorohod Topology (topology of convergence in distribution \cite{MR838085,WLT,LTS}). The main result is expressed in the following theorem.

\begin{thm} Let $J$ be a subset of $I=\{1,\ldots,k\}$, let define for all $j\in J, D_j^\varepsilon=A_j+I/\varepsilon$, where $A_j, j\in J$ are $N\times N$ matrices. Let $N_i(t), i=1,\ldots,k$ be $k$ independent Possion point processes on $\mathbb{R}^2$, let $(W_t=(W_j(t))_{j\in J})$ be a $card(J)$-dimensional Brownian motion independent of the Poisson point processes.

Let assume that the condition
\begin{equation}\label{condition}
\sum_{j\in J}\Big(A_j-A_j^\star\Big)=0
\end{equation}
is satisfied. Therefore, the solution of 
\begin{multline}\label{eqj}
\rho^\varepsilon_t=\int_{0}^t\Big(L(\rho^\varepsilon_{s\mbox{\tiny{-}}})+\sum_{i\in I\setminus J}\left(-D_i\rho^\varepsilon_{s\mbox{\tiny{-}}} D_i^\star+\mathrm{Tr}[D_i\rho^\varepsilon_{s\mbox{\tiny{-}}}D_i^\star]\rho^\varepsilon_{s\mbox{\tiny{-}}}\right)\hfill\\\hphantom{cccccccccccc}+\sum_{j\in  J}\left(-D_j^\varepsilon\rho(D_j^\varepsilon)^\star+\mathrm{Tr}[D_j^\varepsilon\rho^\varepsilon_{s\mbox{\tiny{-}}}(D_j^\varepsilon)^\star]\rho^\varepsilon_{s\mbox{\tiny{-}}}\right)\Big)ds\hfill\\\hphantom{\rho_t=c}+\sum_{i\in I\setminus J}\int_0^t\int_{\mathbb{R}}\left(\frac{D_i\rho^\varepsilon_{s\mbox{\tiny{-}}}D_i^\star}{\mathrm{Tr}\big[D_i\rho^\varepsilon_{s\mbox{\tiny{-}}}D_i^\star\big]}-\rho^\varepsilon_{s\mbox{\tiny{-}}}\right)\mathbf{1}_{0<x<\mathrm{Tr}[D_i\rho^\varepsilon_{s\mbox{\tiny{-}}}D_i^\star]}N_i(dx,ds)\hfill\\
\hphantom{\rho_t=c}+\sum_{j\in J}\int_0^t\int_{\mathbb{R}}\left(\frac{D_j^\varepsilon\rho^\varepsilon_{s\mbox{\tiny{-}}}(D_j^\varepsilon)^\star}{\mathrm{Tr}\big[D_j^\varepsilon\rho^\varepsilon_{s\mbox{\tiny{-}}}(D_j^\varepsilon)^\star\big]}-\rho^\varepsilon_{s\mbox{\tiny{-}}}\right)\mathbf{1}_{0<x<\mathrm{Tr}[D_j^\varepsilon\rho^\varepsilon_{s\mbox{\tiny{-}}}(D_j^\varepsilon)^\star]}N_j(dx,ds),\hfill
\end{multline}
where
$$L_\varepsilon(\rho)=-i[H,\rho]+\sum_{i\in I\setminus J}\left(-\frac{1}{2}\{D_i^\star D_i,\rho\}+D_i\rho D_i^\star\right)+\sum_{j\in  J}\left(-\frac{1}{2}\{(D_j^\varepsilon)^\star D_j^\varepsilon,\rho\}+D_j^\varepsilon\rho(D_j^\varepsilon)^\star\right)$$
converges, when $\varepsilon$ goes to zero, in the space $\mathcal{D}[0,T[$, for all $T$, to the solution of the stochastic differential equation
\begin{multline}\label{eqdif}
	\rho_t=\int_{0}^t\Big(L(\rho_{s\mbox{\tiny{-}}})+\sum_{i\in I\setminus J}(-D_i\rho^\varepsilon_{s\mbox{\tiny{-}}}D_i^\star+\mathrm{Tr}[D_i\rho^\varepsilon_{s\mbox{\tiny{-}}}D_i^\star]\rho^\varepsilon_{s\mbox{\tiny{-}}}\Big)ds\hfill\\\hphantom{\rho_t=c}+\sum_{i\in I\setminus J}\int_0^t\int_{\mathbb{R}}\left(\frac{D_i\rho_{s\mbox{\tiny{-}}}D_i^\star}{\mathrm{Tr}\big[D_i\rho_{s\mbox{\tiny{-}}}D_i^\star\big]}-\rho_{s\mbox{\tiny{-}}}\right)\mathbf{1}_{0<x<\mathrm{Tr}[D_i\rho_{s\mbox{\tiny{-}}}D_i^\star]}N_i(dx,ds)\hfill\\
\hphantom{\rho_t=c}+\sum_{j\in J}\int_0^t\Big(A_j\rho_{s\mbox{\tiny{-}}}+\rho_{s\mbox{\tiny{-}}}A_j^\star-\mathrm{Tr}\big[\rho_{s\mbox{\tiny{-}}}(A_j+A_j^\star)\big]\rho_{s\mbox{\tiny{-}}}\Big)dW_j(s),\hfill
\end{multline}
where
$$L(\rho)=-i[H,\rho]+\sum_{i\in I\setminus J}\left(-\frac{1}{2}\{D_i^\star D_i,\rho\}+D_i\rho D_i^\star\right)+\sum_{j\in  J}\left(-\frac{1}{2}\{A_j^\star A_j,\rho\}+A_j\rho A_j^\star\right).$$
\end{thm} 

It is then straightforward that the above Theorem is a generalization of the heuristic approach presented in the Introduction. The next subsection is devoted to the proof of this result.

\subsection{Proof of Theorem 1}

The proof of Theorem $1$ relies on two steps. First we prove that, under the condition $(\ref{condition})$, the family of processes $(\rho_t^\varepsilon)$ owns the tightness property which corresponds to the relative compactness criterion in the Skorohod Topology. Secondly, we show that the family of Markov generators $(\mathcal{A}_\varepsilon)$ associated with the processes $(\rho_t^\varepsilon)$ converges to the Markov generator of $(\rho_t)$. Next, by combining the tightness property with this result, we get the final convergence.  The tightness property follows from the next proposition.

\begin{pr}\label{tigg} Let $T>0$ be fixed. Let assume that the condition (\ref{condition}) is satisfied. Let $(\rho^{\varepsilon}_t)$ be the solution of the equation (\ref{eqj}). For all $M>0$, there
exists some constant $Z$ such that for all $\varepsilon\leq M$
\begin{equation}\label{tight}
\mathbf{E}\left[\Vert\rho_{t_2}^\varepsilon-\rho_t^\varepsilon\Vert^2\Vert\rho_t^\varepsilon-\rho_{t_1}^\varepsilon\Vert^2\right]\leq Z(t_2-t_1)^2,
\end{equation}
for all $t_1<t<t_2< T$.

 Therefore, the family of processes $(\rho^{\varepsilon}_t)_{0\leq t<T},\varepsilon>0$ is tight for the Skorohod topology on $\mathcal{D}[0,T[$.
\end{pr}

\noindent\begin{pf} The fact that the property $(\ref{tight})$ implies the tightness for the Skorohod topology is a classical result (see Theorem 15.6 in \cite{MR1700749} and Ref \cite{LTS} for further explanations). Here, we just show that the inequality $(\ref{tight})$ is true. Before attacking the estimation, we need to notice the following two facts.

 First, since the condition $(\ref{condition})$ is satisfied, an easy computation gives
\begin{equation}L_\varepsilon(\rho)=L(\rho),\hfill
\end{equation}
for all states $\rho\in\mathcal{S}$. As $\mathcal{S}$ is compact, the function $L_\varepsilon$ is bounded by a constant $K$ on $\mathcal{S}$ independently of $\varepsilon$.

 The second fact concerns the estimation of the terms $D_j^\varepsilon\rho (D_j^\varepsilon)^\star/\textrm{\textrm{Tr}}[D_j^\varepsilon\rho (D_j^\varepsilon)^\star]$, $j\in J$. For all $j\in J$, with the definition of $D_j^\varepsilon$, we get
\begin{equation}\label{estimation1}
\frac{D_j^\varepsilon\rho (D_j^\varepsilon)^\star}{\textrm{\textrm{Tr}}[D_j^\varepsilon\rho (D_j^\varepsilon)^\star]}=\rho+\varepsilon\Big(A_j\rho+\rho A_j^\star-\textrm{Tr}\big[\rho(A_j+A_j^\star)\big]\rho\Big)+\circ(\varepsilon).
\end{equation}
It is important to notice that the $\circ(\varepsilon)$ are uniform on $\mathcal{S}$ (since $\mathcal{S}$ is compact).

 Now, we are in the position to prove the inequality (\ref{tight}). To this end, we introduce the natural filtration $(\mathcal{G}_t^\varepsilon)$ of $(\rho_t^\varepsilon)$, that is, $\mathcal{G}_t^\varepsilon=\sigma\{\rho_u^\varepsilon,u\leq t\}$. In this way, we first estimate the term  $A(t,t_2,\varepsilon)=\mathbf{E}\big[\Vert\rho_{t_2}^\varepsilon-\rho_{t}^\varepsilon\Vert^2\big|\mathcal{G}_t^\varepsilon]$. We have
 \begin{multline}\label{expr11}
 A(t,t_2,\varepsilon)\hfill\\\hphantom{ccccc}\leq\mathbf{E}\left[\left(\int_t^{t_2}\Vert L(\rho_{s\mbox{\tiny{-}}}^\varepsilon)\Vert ds\right)^2\Big|\mathcal{G}_t^\varepsilon\right]\hfill\\\hphantom{ccccc\leq}+\sum_{i\in I\setminus J}\mathbf{E}\left[\left\Vert\int_t^{t_2}\int_\mathbb{R}\left(\frac{D_i\rho^\varepsilon_{s\mbox{\tiny{-}}}D_i^\star}{\textrm{Tr}[D_i\rho^\varepsilon_{s\mbox{\tiny{-}}}D_i^\star]}-\rho_{s\mbox{\tiny{-}}}^\varepsilon\right)\mathbf{1}_{0<x<\textrm{Tr}[D_i\rho^\varepsilon_{s\mbox{\tiny{-}}}D_i^\star]}N_i(dx,ds)\right\Vert^2\Big|\mathcal{G}_t^\varepsilon\right]\hfill\\
 \hphantom{ccccc\leq}+\sum_{j\in J}\mathbf{E}\left[\left\Vert\int_t^{t_2}\int_\mathbb{R}\left(\frac{D_j^\varepsilon\rho^\varepsilon_{s\mbox{\tiny{-}}}(D_j^\varepsilon)^\star}{\textrm{Tr}[D_j^\varepsilon\rho^\varepsilon_{s\mbox{\tiny{-}}}(D_j^\varepsilon)^\star]}-\rho_{s\mbox{\tiny{-}}}^\varepsilon\right)\mathbf{1}_{0<x<\textrm{Tr}[D_j^\varepsilon\rho^\varepsilon_{s\mbox{\tiny{-}}}(D_j^\varepsilon)^\star]}N_j(dx,ds)\right\Vert^2\Big|\mathcal{G}_t^\varepsilon\right].\hfill
 \end{multline}
 Let $(i),(ii),(iii)$ denote the three above terms of the sum.
 
 Since $L$ is bounded on $\mathcal{S}$, for the first term, we have $(i)\leq K(t_2-t)^2\leq KT(t_2-t)$ almost surely.
  
   By remarking that $\textrm{card}\{I\setminus J\}\leq k$, for the term $(ii)$, we have
 \begin{multline}\label{expr12}
 (ii)\leq k\sup_{i\in I\setminus J}\mathbf{E}\left[\int_t^{t_2}\int_\mathbb{R}\left\Vert\left(\frac{D_i\rho^\varepsilon_{s\mbox{\tiny{-}}}D_i^\star}{\textrm{Tr}[D_i\rho^\varepsilon_{s\mbox{\tiny{-}}}D_i^\star]}-\rho_{s\mbox{\tiny{-}}}^\varepsilon\right)\mathbf{1}_{0<\textrm{Tr}[D_i\rho^\varepsilon_{s\mbox{\tiny{-}}}D_i^\star]}\mathbf{1}_{0<x<\textrm{Tr}[D_i\rho^\varepsilon_{s\mbox{\tiny{-}}}D_i^\star]}\right\Vert^2dxds\Big|\mathcal{G}_t^\varepsilon\right]\hfill\\
 \hphantom{(ii)\leq}\leq k\sup_{i\in I\setminus J}\mathbf{E}\left[\int_t^{t_2}\left\Vert\left(\frac{D_i\rho^\varepsilon_{s\mbox{\tiny{-}}}D_i^\star}{\textrm{Tr}[D_i\rho^\varepsilon_{s\mbox{\tiny{-}}}D_i^\star]}-\rho_{s\mbox{\tiny{-}}}^\varepsilon\right)\mathbf{1}_{0<\textrm{Tr}[D_i\rho^\varepsilon_{s\mbox{\tiny{-}}}D_i^\star]}\right\Vert^2 \textrm{Tr}[D_i\rho^\varepsilon_{s\mbox{\tiny{-}}}D_i^\star]ds\Big|\mathcal{G}_t^\varepsilon\right].  \hfill
 \end{multline}
 The passage from the expression $(ii)$ in (\ref{expr11}) to the first inequality in (\ref{expr12}) relies on two facts. First, we have computed the expectation with respect to the Poisson point processes by using the usual property of stochastic integration theory with respect to random measures (such a property is called Ito isometry property). In other words we have used that the intensity measure of each Poison point process $N_i$ is the Lebesgue measure (we refer to \cite{CSPM} for a complete introduction of stochastic integration theory with respect to random measure). Secondly, we have introduced the term $\mathbf{1}_{0<\textrm{Tr}[D_i\rho^\varepsilon_{s\mbox{\tiny{-}}}D_i^\star]}$ by remarking that $\mathbf{1}_{0<x<\textrm{Tr}[D_i\rho^\varepsilon_{s\mbox{\tiny{-}}}D_i^\star]}=\mathbf{1}_{0<\textrm{Tr}[D_i\rho^\varepsilon_{s\mbox{\tiny{-}}}D_i^\star]}\mathbf{1}_{0<x<\textrm{Tr}[D_i\rho^\varepsilon_{s\mbox{\tiny{-}}}D_i^\star]}$ (this allows to keep the fact that we have the property $0<\textrm{Tr}[D_i\rho^\varepsilon_{s\mbox{\tiny{-}}}D_i^\star]$ in the second inequality). Thus, with respect to the underlying probability, it is important to notice that we implicitly consider that the quantity $D_i\rho^\varepsilon_{s\mbox{\tiny{-}}}D_i^\star/\textrm{Tr}[D_i\rho^\varepsilon_{s\mbox{\tiny{-}}}D_i^\star]=0$ if $\textrm{Tr}[D_i\rho^\varepsilon_{s\mbox{\tiny{-}}}D_i^\star]=0$, otherwise this quantity defines a state. Now, since the set of states $\mathcal{S}$ is compact, the term inside the $L_2$ norm is almost surely bounded, then the term inside the integral $\int_t^{t_2}$ is bounded. As a consequence, there exists a constant $K_1$ such that $(ii)\leq K_1(t_2-t)$ almost surely. 

 For the term $(iii)$, since $\textrm{card}\{ J\}\leq k$, by using the estimation $(\ref{estimation1})$, we have
 \begin{multline}
 (iii)\leq k\sup_{j\in  J}\mathbf{E}\left[\int_t^{t_2}\int_\mathbb{R}\left\Vert\left(\frac{D_j^\varepsilon\rho^\varepsilon_{s\mbox{\tiny{-}}}(D_j^\varepsilon)^\star}{\textrm{Tr}[D_j^\varepsilon\rho^\varepsilon_{s\mbox{\tiny{-}}}(D_j^\varepsilon)^\star]}-\rho_{s\mbox{\tiny{-}}}^\varepsilon\right)\mathbf{1}_{0<x<\textrm{Tr}[D_j^\varepsilon\rho^\varepsilon_{s\mbox{\tiny{-}}}(D_j^\varepsilon)^\star]}\right\Vert^2dxds\Big|\mathcal{G}_t^\varepsilon\right]\hfill\\
 \hphantom{(iii)\leq}\leq k\sup_{j\in  J}\mathbf{E}\left[\int_t^{t_2}\left\Vert\left(\frac{D_j^\varepsilon\rho^\varepsilon_{s\mbox{\tiny{-}}}(D_j^\varepsilon)^\star}{\textrm{Tr}[D_j^\varepsilon\rho^\varepsilon_{s\mbox{\tiny{-}}}(D_j^\varepsilon)^\star]}-\rho_{s\mbox{\tiny{-}}}^\varepsilon\right)\right\Vert^2\textrm{Tr}[D_j^\varepsilon\rho^\varepsilon_{s\mbox{\tiny{-}}}(D_j^\varepsilon)^\star]ds\Big|\mathcal{G}_t^\varepsilon\right]\hfill\\
 \hphantom{(iii)\leq}\leq k\sup_{j\in  J}\mathbf{E}\Bigg[\int_t^{t_2}\left\Vert\varepsilon\Big(A_j\rho^\varepsilon_{s\mbox{\tiny{-}}}+\rho^\varepsilon_{s\mbox{\tiny{-}}} A_j^\star-\textrm{Tr}\big[\rho^\varepsilon_{s\mbox{\tiny{-}}}(A_j+A_j^\star)\big]\rho^\varepsilon_{s\mbox{\tiny{-}}}\Big)+\circ(\varepsilon)\right\Vert^2\hfill\\\hphantom{cccccccccccccccccccccc}\times\left(\frac{1}{\varepsilon^2}+\frac{1}{\varepsilon}\textrm{Tr}[\rho^\varepsilon_{s\mbox{\tiny{-}}}(A_j+A_j^\star)]+\textrm{Tr}[A_j\rho_{s\mbox{\tiny{-}}}^\varepsilon A_j^\star]\right)ds\Big|\mathcal{G}_t^\varepsilon\Bigg].\hfill
  \end{multline}
  
 Now, it is straightforward to notice that the factor $\varepsilon$, in the $L_2$ norm, compensates the terms $1/\varepsilon^2$ and $1/\varepsilon$. Hence, for $M$ being fixed, there exists a constant $K_2$ such that for all $\varepsilon\leq M$, the term inside the expectation is bounded (recall that $\mathcal{S}$ is compact). Then, we have $(iii)\leq K_2(t_2-t)$ almost surely. As a consequence, for all $M>0$, there exists a constant $S=KT+K_1+K_2$ independent of $t,t_2$ and $\varepsilon$, such that for all $\varepsilon\leq M$, we have almost surely
 $$A(t,t_2,\varepsilon)\leq S(t_2-t).$$
 We shall show that this implies the expected result. By conditioning with $\mathcal{G}_t^\varepsilon$, we have
 \begin{eqnarray}
 \mathbf{E}\left[\Vert\rho_{t_2}^\varepsilon-\rho_t^\varepsilon\Vert^2\Vert\rho_t^\varepsilon-\rho_{t_1}^\varepsilon\Vert^2\right]&=&\mathbf{E}\left[A(t,t_2,\varepsilon)\Vert\rho_t^\varepsilon-\rho_{t_1}^\varepsilon\Vert^2\right]\nonumber\\
 &\leq&S(t_2-t)\mathbf{E}\left[A(t_1,t,\varepsilon)\right]\nonumber\\
 &\leq&S^2(t_2-t)(t-t_1)\nonumber\\
 &\leq&\frac{S^2}{2}(t_2-t_1)^2
 \end{eqnarray}
 and the result is proved with $Z=S^2/2$.
\end{pf}
\bigskip

Now, we address the second step which concerns the convergence of Markov generators. Moreover, this convergence gives a reverse result concerning the condition $(\ref{condition})$. Before expressing the result, we introduce the notation $C^2_c$ for denoting the set of $C^2$ functions defined on $\mathcal{B}(\mathcal{H})$ with values in $\mathbb{R}$ and with compact support. The terms, $D_\rho f(.)$ and $D_\rho^2f(.,.)$ will denote the first and second differential. Besides, we define the functions $h_j,j\in J$ on $\mathcal{S}$ by
$$h_j(\rho)=A_j\rho+\rho A_j^\star-\mathrm{Tr}\big[\rho(A_j+A_j^\star)\big]\rho,$$
for all states $\rho$.
\begin{pr}
Let $\mathcal{A_\varepsilon}$ be the Markov generator of the process $(\rho_t^\varepsilon)$ defined by
\begin{eqnarray}
\mathcal{A_\varepsilon}f(\rho)&=&D_\rho f\big(L_\varepsilon(\rho)\big)\nonumber\\&&+\sum_{i\in I\setminus J}\left[ f\left(\frac{D_i\rho D_i^\star}{\mathrm{Tr}[D_i\rho D_i^\star]}\right)-f(\rho)-D_\rho f\left(\frac{D_i\rho D_i^\star}{\mathrm{Tr}[D_i\rho D_i^\star]}-\rho\right)\right]\mathrm{Tr}[D_i\rho D_i^\star]\nonumber\\&&+\sum_{j\in J}\left[f\left(\frac{D_j^\varepsilon\rho (D_j^\varepsilon)^\star}{\mathrm{Tr}[D_j^\varepsilon\rho (D_j^\varepsilon)^\star]}\right)-f(\rho)-D_\rho f\left(\frac{D_j^\varepsilon\rho (D_j^\varepsilon)^\star}{\mathrm{Tr}[D_j^\varepsilon\rho (D_j^\varepsilon)^\star]}-\rho\right)\right]\mathrm{Tr}[D_j^\varepsilon\rho (D_j^\varepsilon)^\star], \nonumber\\&&
\end{eqnarray}
for all $f\in C^2_c$ and for all states $\rho$.

 Let $\mathcal{A}$ be the Markov generator of the process $(\rho_t)$ defined by
\begin{eqnarray}
\mathcal{A}f(\rho)&=&D_\rho f\big(L(\rho)\big)\nonumber\\&&+\sum_{i\in I\setminus J}\left[ f\left(\frac{D_i\rho D_i^\star}{\mathrm{Tr}[D_i\rho D_i^\star]}\right)-f(\rho)-D_\rho f\left(\frac{D_i\rho D_i^\star}{\mathrm{Tr}[D_i\rho D_i^\star]}-\rho\right)\right]\mathrm{Tr}[D_i\rho D_i^\star]\nonumber\\&&+\frac{1}{2}\sum_{j\in J}D_\rho^2f\big(h_j(\rho),h_j(\rho)\big),
\end{eqnarray}
for all $f\in C^2_c$ and for all states $\rho$.

 Then, let $f\in C^2_c$, we have the convergence
\begin{equation}
\lim_{\varepsilon\rightarrow 0}\,\,\sup_{\rho\in\mathcal{S}}\,\,\vert\mathcal{A_\varepsilon}f(\rho)-\mathcal{A}f(\rho)\vert=0,
\end{equation} 
 if and only if the condition $(\ref{condition})$:
$
\sum_{j\in J}\Big(A_j-A_j^\star\Big)=0
$ is satisfied.

As a consequence, we have
\begin{equation}\label{CC}
 \lim_{\varepsilon\rightarrow0}\mathbf{E}\left[\left(f(\rho^\varepsilon_{t+s})-f(\rho^\varepsilon_t)
 -\int_t^{t+s}\mathcal{A}f(\rho^\varepsilon_s)ds\right)\prod_{i=1}^m\theta_i(\rho^\varepsilon_{t_i})\right]=0,
\end{equation}
for all $m\geq0$, for all $0\leq t_1<t_2<\ldots<t_m\leq t<t+s$, for all functions
 $(\theta_i)_{i=1,\ldots, m}$ and for all $f$ in $C^2_c$.
\end{pr}

\noindent\textbf{Remark:} The fact that the operators $\mathcal{A}_\varepsilon$ and $\mathcal{A}$ are the Markov generators of $(\rho_t^\varepsilon)$ and $(\rho_t)$ follows from Ito stochastic calculus (see \cite{MR838085,CSPM} concerning the definition of Markov generators and \cite{p3} for explicit computations). Moreover, we have the following important property. 

Let $\rho_0$ be a state and let $f\in C^2_c$, the process defined by
\begin{equation}\label{martprob}f(\rho_{t}^\varepsilon)-f(\rho_0)-\int_0^t\mathcal{A}_\varepsilon f(\rho_s^\varepsilon)ds,\end{equation}
for all $t \geq 0$, is a martingale with respect to the natural filtration  $(\mathcal{G}_t^\varepsilon)$ of $(\rho_t^\varepsilon)$, where $\mathcal{G}_t^\varepsilon=\sigma\{\rho_u^\varepsilon, u\leq t\}$. This property is related to the notion of \textit{Martingale problem}. Let us make precise this notion. In probability theory, the couple $(\mathcal{A}_\varepsilon,\rho_0)$ defines what is called a \textit{Martingale problem}. Solving this martingale problem consists in finding a Markov process $(X_t)$ such that $X_0=\rho_0$ and such that, for all $f\in C^2_c$, the process $(f(X_t)-f(X_0)-\int_0^t\mathcal{A}_\varepsilon f(X_s)ds)$ is a martingale with respect to the natural filtration generated by $(X_t)$ (see \cite{CSPM,MR838085,LTS} for more general definitions of martingale problem). In this sense the process $(\rho_{t}^\varepsilon)$ is a solution of the martingale problem $(\mathcal{A}_\varepsilon,\rho_0)$ (the same holds for $(\rho_t)$ with $(\mathcal{A},\rho_0)$).

 Moreover, as $(\rho_t^\varepsilon)$ is the unique solution of the equation $(\ref{eqj})$ and as the property $(\ref{martprob})$ is satisfied, the process $(\rho_t^\varepsilon)$ is the unique solution, \textbf{in law}, of the martingale problem $(\mathcal{A}_\varepsilon,\rho_0)$ (see Refs. \cite{MR838085,LTS,QR,CSPM} for general considerations on martingale problems and \cite{p3} for results concerning stochastic master equations). Note that the process $(\rho_t)$ is also the unique solution, in law, of the martingale problem associated with $(\mathcal{A},\rho_0)$. 
 
 Since we shall prove a result of convergence in distribution, the result of uniqueness, in law, is crucial. Indeed, if the limit process of $(\rho_t^\varepsilon)$ satisfies the martingale property $(\ref{martprob})$ for $(\mathcal{A},\rho_0)$, this limit process will be equal to $(\rho_t)$ in law. Let us prove, now, the proposition.

\noindent\begin{pf} Let $j\in J$, the estimation (\ref{estimation1}) stated in the proof of Proposition \ref{tigg}, allows to apply the Taylor formula. This gives\begin{multline}\left[f\left(\frac{D_j^\varepsilon\rho (D_j^\varepsilon)^\star}{\textrm{\textrm{Tr}}[D_j^\varepsilon\rho (D_j^\varepsilon)^\star]}\right)-f(\rho)-D_\rho f\left(\frac{D_j^\varepsilon\rho (D_j^\varepsilon)^\star}{\textrm{\textrm{Tr}}[D_j^\varepsilon\rho (D_j^\varepsilon)^\star]}-\rho\right)\right]\textrm{\textrm{Tr}}[D_j^\varepsilon\rho (D_j^\varepsilon)^\star]=\hfill\\\hphantom{cccccccccccccccccccccccccccccccccccccccc}=\frac{1}{2}D_\rho^2f\big(h_j(\rho), h_j(\rho)\big)+\circ(\varepsilon),\hfill\end{multline}
where $\circ(\varepsilon)$ is uniform in $\rho$. As a consequence, it is easy to see that
\begin{multline}\label{con1}
\lim_{\varepsilon\rightarrow 0}\sup_{\rho\in\mathcal{S}}\Bigg\vert\Bigg(\sum_{j\in J}\left[f\left(\frac{D_j^\varepsilon\rho (D_j^\varepsilon)^\star}{\textrm{\textrm{Tr}}[D_j^\varepsilon\rho (D_j^\varepsilon)^\star]}\right)-f(\rho)-D_\rho f\left(\frac{D_j^\varepsilon\rho (D_j^\varepsilon)^\star}{\textrm{\textrm{Tr}}[D_j^\varepsilon\rho (D_j^\varepsilon)^\star]}-\rho\right)\right]\textrm{\textrm{Tr}}[D_j^\varepsilon\rho (D_j^\varepsilon)^\star]\hfill\\\hphantom{cccccccccccccccccc}-\frac{1}{2}D_\rho^2\big(h_j(\rho), h_j(\rho)\big)\Bigg)\Bigg\vert=0. \hfill
\end{multline} This first approximation shows how the diffusive part, in terms of Markov generators, appears in the limit. Let us treat now the term $D_\rho f(L_\varepsilon)$. The only contribution involves the terms where the operators $D_j^\varepsilon$ appear. For all $j\in J$, we have
\begin{multline}\label{conv2}
D_\rho f\left(-\frac{1}{2}\{(D_j^\varepsilon)^\star D_j^\varepsilon,\rho\}+D_j^\varepsilon\rho(D_j^\varepsilon)^\star\right)=\hfill\\
\hphantom{D_\rho f\left(-\frac{1}{2})^\star D_j^\varepsilon\right.}=D_\rho f\left(-\frac{1}{2}\{A_j^\star A_j,\rho\}+A_j\rho A_j^\star\right)+\frac{1}{2\varepsilon}D_\rho f\left(\big(A_j-A_j^\star\big)\rho+\rho\big(A_j-A_j^\star\big)\right).\hfill
\end{multline}
As a consequence, the two expressions (\ref{con1}) and (\ref{conv2}) imply the equivalence
\begin{equation}
\lim_{\varepsilon\rightarrow 0}\,\,\sup_{\rho\in\mathcal{S}}\,\,\vert\mathcal{A_\varepsilon}f(\rho)-\mathcal{A}f(\rho)\vert=0, \forall f\in C_c^2\,\,\Leftrightarrow\,\,\sum_{j\in J}\Big(A_j-A_j^\star\Big)\rho+\rho\Big(A_j-A_j^\star\Big)=0, \forall\rho\in \mathcal{S}.
\end{equation} 
Now, if we have $\sum_{j\in J}\Big((A_j-A_j^\star)\rho+\rho(A_j-A_j^\star)\Big)=0,$ for all states, this equality holds for all positive matrices (by multiplying by the trace). Since any Hermitian matrix can be written as linear combination of positive matrices (this comes from the diagonalization), this equality holds for all Hermitian matrices. Now, since we have $X=(X+X^\star)/2+i(-iX+iX^\star)/2$, for all matrices $X$ and since the matrices $X+X^\star$ and $-iX+iX^\star$ are Hermitian, the equality holds for all matrices. Thus, we get the complete characterization  $(\ref{condition})$ and the equivalence is proved (actually the uniform convergence is not necessary to imply the condition (\ref{condition})).

 In order to finish the proof of the proposition it remains to prove the convergence (\ref{CC}). To this end, we insert the term $\int_t^{t+s}\mathcal{A}_\varepsilon f(\rho^\varepsilon_u)du$ in the expectation. Next, according to the remark before the proof, we use the fact that $f(\rho_t^\varepsilon)-f(\rho_0)-\int_0^t\mathcal{A}_\varepsilon f(\rho^\varepsilon_u)du$ is a martingale with respect to the natural filtration $(\mathcal{G}_t^\varepsilon)$ of $(\rho_t^\varepsilon)$ (we have to notice that $\prod_{i=1}^m\theta_i(\rho^\varepsilon_{t_i})$ is $\mathcal{G}_t^\varepsilon$ measurable). Next, the uniform convergence result of Markov generators ensures the convergence $(\ref{CC})$.
\end{pf} 
\bigskip
 
 Now, we are in the position to prove Theorem 1. Let us assume that the condition (\ref{condition}) is satisfied. Let $(\varepsilon_n)$ be a sequence converging to $0$. Since $(\rho_t^{\varepsilon_n})$ is a tight sequence (Proposition 1), we can extract a convergence subsequence. Let denote $(\mu_t)$ the limit. According to the convergence $(\ref{CC})$, the process $(\mu_t)$ satisfies 
$$\mathbf{E}\left[\left(f(\mu_{t+s})-f(\mu_t)
 -\int_t^{t+s}\mathcal{A}f(\mu_s)ds\right)\prod_{i=1}^m\theta_i(\mu_{t_i})\right]=0,$$
for all $m\geq0$, for all $0\leq t_1<t_2<\ldots<t_m\leq t<t+s$, for all functions
 $(\theta_i)_{i=1,\ldots, m}$ and for all $f$ in $C^2_c$. This implies that $(\mu_t)$ is the solution of the problem of martingale associated with $\mathcal{A}$. By uniqueness, in law, of the solution of the martingale problem, the process $(\mu_t)$ has the same distribution than $(\rho_t)$. In this way, every convergent sequence of processes $(\rho_t^{\varepsilon_n})$, where $(\varepsilon_n)$ converges to $0$, converges in distribution to $(\rho_t)$. As a conclusion, the family of processes $(\rho_t^\varepsilon)$ converges in distribution to $(\rho_t)$. 
 
 \section{Applications}

This section is devoted to present situations where the condition $(\ref{condition})$ is naturally satisfied and where the diffusion approximation can be performed. 
As a first direct application, we see that the heuristic approach presented in the Introduction is rigorously justified by Theorem 1. Indeed the condition $C=C^\star$ is no more than the condition (\ref{condition}).
In the two following subsections, we investigate the diffusion approximation in the model of homodyne and heterodyne detection.

\subsection{Diffusion Approximation in Homodyne Setup}

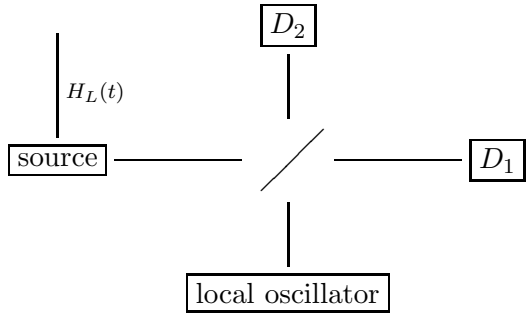
\begin{figure}
\caption{Experimental setup of homodyning.}
\label{Fig1}
\begin{equation*}
    \xymatrix{%
     \ar@{-}[d]^{H_{L}(t)} & \text{\framebox{$D_2$}} \\
      \text{\framebox{\text{source}}} \ar@{-}[r] & \thickspace
     \begin{turn}{135}
        \Bigg{\vert}
      \end{turn}
      \ar@{-}[u]
      \ar@{-}[r] \ar@{-}[d] & \text{\framebox{$D_1$}} \\
      & \text{\framebox{\text{local oscillator}}}
      }
	\end{equation*}
	\end{figure}

The model of homodyne detection describes a two-level atom whose emitted light is detected by photodetection. The source, i.e, the two level atom, is driven by an external interaction. The light, emitted by the atom traverses a beam splitter and interferes with the light field of a local oscillator. Then, two detectors $D_1$ and $D_2$ detect the light (see Section 6.4 of \cite{francesco2} or \cite{Book,W1,W22,W2,carm1,carm2} for more details on the experimental setup and optical considerations, see Figure \ref{Fig1}). Typically, we record two different types of jumps according of which detectors detects the light.

Our aim is not to derive the equation describing this situation, we refer to \cite{BAR,Book,MR1639788,SSEMB,francesco2,milb,carm1,carm2,Gisin2,W1,W22,W2} for justifications. Let us give the stochastic master equations governing the evolution of the two level atom. To this end, we consider the stochastic Schr\"odinger equation, that is, the stochastic differential equation which describes the evolution of the two-level atom in terms of wave functions (or pure states). Here, the corresponding equation is given by
\begin{eqnarray}\label{purejump} 
\psi_t&=&\psi_0+\int_0^t-i\Big(\hat{H}+\frac{i\gamma_0}{2}\Vert\sigma_{\mbox{\scriptsize{-}}}\psi_{s\mbox{\tiny{-}}}\Vert^2\Big)\psi_{s\mbox{\tiny{-}}}ds\nonumber\\
&&+\int_0^t\int_\mathbb{R}\left(\frac{(\sigma_{\mbox{\scriptsize{-}}}+i\beta)\psi_{s\mbox{\tiny{-}}}}{\Vert(\sigma_{\mbox{\scriptsize{-}}}+i\beta)\psi_{s\mbox{\tiny{-}}}\Vert}-\psi_{s\mbox{\tiny{-}}}\right)\mathbf{1}_{0<x<\frac{\gamma_0}{2}\Vert(\sigma_{\mbox{\scriptsize{-}}}+i\beta)\psi_{s\mbox{\tiny{-}}}\Vert^2}N_1(dx,ds)\nonumber\\
&&+\int_0^t\int_\mathbb{R}\left(\frac{(\sigma_{\mbox{\scriptsize{-}}}-i\beta)\psi_{s\mbox{\tiny{-}}}}{\Vert(\sigma_{\mbox{\scriptsize{-}}}-i\beta)\psi_{s\mbox{\tiny{-}}}\Vert}-\psi_{s\mbox{\tiny{-}}}\right)\mathbf{1}_{0<x<\frac{\gamma_0}{2}\Vert(\sigma_{\mbox{\tiny{-}}}-i\beta)\psi_{s\mbox{\tiny{-}}}\Vert^2}N_2(dx,ds),
\end{eqnarray}
where $N_1$ and $N_2$ are two Poisson point processes on $\mathbb{R}^2$.
\bigskip

\noindent\textbf{Remark:} In this equation, the constant $\gamma_0$ represents the \textit{spontaneous emission rate}. As we can see in the equation (\ref{purejump}), this parameter appears in the stochastic intensity. The quantity $\beta$ represents the \textit{amplitude} of the local oscillator. The operator $\hat{H}$ is the free Hamiltonian of the two-level atom and $\sigma_{\mbox{\scriptsize{-}}}=\left(\begin{array}{cc}0&1\\0&0\end{array}\right)$ is the usual notation for the corresponding Pauli matrix.
\bigskip

The corresponding equation for $(\rho_t)$, in terms of density matrices is obtained by defining $\rho_t=\vert\psi_t\rangle\langle\psi_t\vert$ (the notation $\vert\psi_t\rangle\langle\psi_t\vert$ corresponds to the Dirac notation for the one dimensional projector on $\mathbb{C}\psi_t$; one can find also the term \textit{pure state}\footnote{A pure state is a state which is a one-dimensional projector.} to define such kind of state). The equation for $(\rho_t)$ can be derived using the Ito rules (see \cite{Book,barchielli0,p1,p2} for computations). We get then the following stochastic master equation
\begin{eqnarray}
\rho_t&=&\rho_0+\int_0^t\Big(-i[\hat{H},\rho_{s\mbox{\tiny{-}}}]-\frac{1}{2}\{(\sigma_{\mbox{\scriptsize{-}}}+i\beta)^\star(\sigma_{\mbox{\scriptsize{-}}}+i\beta),\rho_{s\mbox{\tiny{-}}}\}-\frac{1}{2}\{(\sigma_{\mbox{\scriptsize{-}}}-i\beta)^\star(\sigma_{\mbox{\scriptsize{-}}}-i\beta),\rho_{s\mbox{\tiny{-}}}\}\nonumber\\
&&\hphantom{cccccccccc}+\textrm{Tr}[(\sigma_{\mbox{\scriptsize{-}}}+i\beta)\rho_{s\mbox{\tiny{-}}}(\sigma_{\mbox{\scriptsize{-}}}+i\beta)^\star]\rho_{s\mbox{\tiny{-}}}+\textrm{Tr}[(\sigma_{\mbox{\scriptsize{-}}}-i\beta)\rho_{s\mbox{\tiny{-}}}(\sigma_{\mbox{\scriptsize{-}}}-i\beta)^\star]\rho_{s\mbox{\tiny{-}}}\Big)ds
\nonumber\\
&&+\int_0^t\int_\mathbb{R}\left(\frac{(\sigma_{\mbox{\scriptsize{-}}}+i\beta)\rho_{s\mbox{\tiny{-}}}(\sigma_{\mbox{\scriptsize{-}}}+i\beta)^\star}{\textrm{Tr}[(\sigma_{\mbox{\scriptsize{-}}}+i\beta)\rho_{s\mbox{\tiny{-}}}(\sigma_{\mbox{\scriptsize{-}}}+i\beta)^\star]}-\rho_{s\mbox{\tiny{-}}}\right)\mathbf{1}_{0<x<\frac{\gamma_0}{2}\textrm{Tr}[(\sigma_{\mbox{\scriptsize{-}}}+i\beta)\rho_{s\mbox{\tiny{-}}}(\sigma_{\mbox{\scriptsize{-}}}+i\beta)^\star]}N_1(dx,ds)\nonumber\\
&&+\int_0^t\int_\mathbb{R}\left(\frac{(\sigma_{\mbox{\scriptsize{-}}}-i\beta)\rho_{s\mbox{\tiny{-}}}(\sigma_{\mbox{\scriptsize{-}}}-i\beta)^\star}{\textrm{Tr}[(\sigma_{\mbox{\scriptsize{-}}}-i\beta)\rho_{s\mbox{\tiny{-}}}(\sigma_{\mbox{\scriptsize{-}}}-i\beta)^\star]}-\rho_{s\mbox{\tiny{-}}}\right)\mathbf{1}_{0<x<\frac{\gamma_0}{2}\textrm{Tr}[(\sigma_{\mbox{\scriptsize{-}}}-i\beta)\rho_{s\mbox{\tiny{-}}}(\sigma_{\mbox{\scriptsize{-}}}-i\beta)^\star]}N_2(dx,ds)\nonumber.
\end{eqnarray}

Now, in order to address a diffusion approximation setup, we denote
$$\beta=i\vert\beta\vert e^{i\theta}$$
and we investigate the limit $\varepsilon=\sqrt{2}/(\sqrt{\gamma_0}\vert\beta\vert)\rightarrow0.$ Physically, such a limit corresponds to a strongly excited local oscillator ($\vert\beta\vert\rightarrow\infty$). Naturally, here, we define
$$D_1^\varepsilon=-\sqrt{\frac{\gamma_0}{2}}\sigma_{\mbox{\scriptsize{-}}}e^{-i\theta}+\frac{I}{\varepsilon}\,\,\,\,\textrm{and}\,\,\,\,D_2^\varepsilon=\sqrt{\frac{\gamma_0}{2}}\sigma_{\mbox{\scriptsize{-}}}e^{-i\theta}+\frac{I}{\varepsilon}.$$
It is then straightforward that $$\sqrt{\frac{\gamma_0}{2}}\Big(-\sigma_{\mbox{\scriptsize{-}}}e^{-i\theta}-(-\sigma_{\mbox{\scriptsize{-}}}e^{-i\theta})^\star+\sigma_{\mbox{\scriptsize{-}}}e^{-i\theta}-(\sigma_{\mbox{\scriptsize{-}}}e^{-i\theta})^\star\Big)=0.$$
This corresponds to the condition (\ref{condition}). Applying Theorem 1 and putting $C=\sigma_{\mbox{\scriptsize{-}}}e^{-i\theta}$, we obtain the diffusion equation
\begin{eqnarray}\label{diffff}
\rho_t&=&\rho_0+\int_{0}^t\Big(-i[\hat{H},\rho_{s}]-\frac{\gamma_0}{4}\{C^\star C,\rho_s\}-\frac{\gamma_0}{4}\{C^\star C,\rho_s\}+\frac{\gamma_0}{2}C\rho_s C^\star+\frac{\gamma_0}{2}C\rho_s C^\star\Big)ds\nonumber\\&&+\int_0^t-\sqrt{\frac{\gamma_0}{2}}\Big(C\rho_s+\rho_sC^\star-\textrm{Tr}[\rho_s(C+C^\star)]\rho_s\Big)dW_1(t)\nonumber\\&&+\int_0^t\sqrt{\frac{\gamma_0}{2}}\Big(C\rho_s+\rho_sC^\star-\textrm{Tr}[\rho_s(C+C^\star)]\rho_s\Big)dW_2(t),
\end{eqnarray}
where $((W_1(t),W_2(t))$ is a $2$-dimensional Brownian motion. Let us notice that we can define $W_t=\sqrt{\frac{1}{2}}\big(W_2(t)-W_1(t)\big)$, which is a standard Brownian motion. The equation (\ref{diffff}) becomes then
\begin{equation}\label{difg}
\rho_t=\rho_0+\int_0^tL(\rho_s)ds+\int_0^t\sqrt{\gamma_0}\Big(C\rho_s+\rho_sC^\star-\textrm{Tr}[\rho_s(C+C^\star)]\rho_s\Big)dW_t,
\end{equation}
where
$$L(\rho)=-i[\hat{H},\rho]-\frac{\gamma_0}{2}\{C^\star C,\rho_s\}+\gamma_0C\rho_s C^\star.$$
We recover the equation which has been presented, in the Introduction, as the model of homodyne detection. 

\subsection{Diffusion Approximation in Heterodyne Setup}

Our last application is the model of heterodyne detection. In this case the parameter $\beta$ is replaced by $\beta\leftrightarrow\beta_t=\beta e^{-i\Delta t},$ where $\Delta$ is the \textit{detuning} of the local oscillator. By assuming that the result of Theorem $1$ is still valid for coefficients depending on time (there are no additional difficulties and the proof is the same). We get a similar expression for the heterodyne detection. In this case the operator $C$ is replaced by a time dependent operator
$C(t)=Ce^{i\Delta_t}.$ Physically, a natural assumption is $\Delta\gg1,$ that is $\Delta\rightarrow\infty$. Let us investigate how the equation $(\ref{difg})$ is transformed under this condition. To this end, let us introduce some elements. We consider the following generators, defined for all $f\in C_c^2$, all $\rho\in\mathcal{S}$ and all $s\in\mathbb{R}$ by
\begin{eqnarray}
\mathcal{A}_{\Delta}f(s,\rho)&=&D_\rho f(L(\rho))+\frac{1}{2}D_\rho^2f(h(s,\Delta,\rho),h(s,\Delta,\rho)),\nonumber\\
\mathcal{A}f(\rho)&=&D_\rho f(L(\rho))+\frac{1}{2}D_\rho^2f(h_+(\rho),h_+(\rho))+\frac{1}{2}D_\rho^2f(ih_-(\rho),ih_-(\rho)),
\end{eqnarray}
where
\begin{eqnarray}h(s,\Delta,\rho)&=&\sqrt{\gamma_0}\Big(C(s)\rho+\rho C(s)^\star-\textrm{Tr}[\rho(C(s)+C(s)^\star)]\rho\Big),\\
h_{\pm}(\rho)&=&\sqrt{\frac{\gamma_0}{2}}\Big(C\rho\pm\rho C^\star-\textrm{Tr}[\rho(C\pm C^\star)]\rho\Big).
\end{eqnarray}
Let $\rho_0$ be a state and let $(\rho_t^\Delta)$ and $(\rho_t)$ be the solutions of the problems of martingale associated with $(\mathcal{A}_\Delta,\rho_0)$ and $(\mathcal{A},\rho_0)$. These solutions can be expressed as solutions of the following stochastic differential equations
\begin{eqnarray}\label{eq1}
\rho_t^\Delta&=&\rho_0+\int_0^tL(\rho^\Delta_s)ds+\int_0^th(s,\Delta,\rho^\Delta_s)dW_s\\\label{eq2}
\rho_t&=&\rho_0+\int_0^tL(\rho_s)ds+\int_0^th_+(\rho_s)dW_1(s)+\int_0^tih_-(\rho_s)dW_2(s),
\end{eqnarray}
where $(W_t)$, $(W_1(t))$, and $(W_2(t))$ are three independent Brownian motions, defined on a same probability space $(\Omega,\mathcal{F},P)$. Now, we are in the position to express the limit result.

\begin{pr}
Let $(\rho_t^\Delta)$ be the solution of (\ref{eq1}) and let $(\rho_t)$ be the solution of $(\ref{eq2})$. 

Therefore, the family of processes $(\rho_t^\Delta)$  converges in distribution to the process $(\rho_t)$, when $\Delta$ goes to infinity.
\end{pr}

\noindent\begin{pf} We will not show the tightness property of the family $(\rho_t^\Delta)$ (one can show a similar result as Proposition $1$ in the previous Section). Here, we cannot use of the convergence of generators as in Proposition $2$. Actually, the convergence of $\mathcal{A}_\Delta$ to $\mathcal{A}$ is not true since $e^{i\Delta t}$ has no limit, when $\Delta$ goes to infinity. Here, we show directly that for all sequences $\Delta_n$, which converge to infinity such that $(\rho_t^{\Delta_n})$ converges in distribution, we have
\begin{equation}\label{CC1}
 \lim_{n\rightarrow\infty}\mathbf{E}\left[\left(f(\rho^{\Delta_n}_{t+s})-f(\rho^{\Delta_n}_t)
 -\int_t^{t+s}\mathcal{A} f(\rho^{\Delta_n}_u)du\right)\prod_{i=1}^m\theta_i(\rho^{\Delta_n}_{t_i})\right]=0,
\end{equation}
for all $m\geq0$, for all $0\leq t_1<t_2<\ldots<t_m\leq t<t+s$, for all functions
 $(\theta_i)_{i=1,\ldots, m}$ and for all $f$ in $C^2_c$ (let us remind that in the previous section, this result was just a consequence of the convergence of generators).  Let $(\Delta_n)$ be such a sequence and let $(\mu_t)$ be the limit process. Since $(\rho_t^{\Delta_n})$ is valued is in the set of states, we first remark that the limit process $(\mu_t)$ is also valued in the set of states (the property satisfied by a state are closed for the topology of Skorohod). Now, we can remark \begin{multline}\label{35}
 \mathbf{E}\left[\left(f(\rho^{\Delta_n}_{t+s})-f(\rho^{\Delta_n}_t)
 -\int_t^{t+s}\mathcal{A} f(\rho^{\Delta_n}_u)du\right)\prod_{i=1}^m\theta_i(\rho^{\Delta_n}_{t_i})\right]=\hfill\\
 \hphantom{ccccccccccccc}=\mathbf{E}\left[\left(\int_t^{t+s}\mathcal{A}_{\Delta_n} f(u,\rho^{\Delta_n}_u)du-\int_t^{t+s}\mathcal{A} f(\rho^{\Delta_n}_u)du\right)\prod_{i=1}^m\theta_i(\rho^{\Delta_n}_{t_i})\right]\hfill\\
 \hphantom{ccccccccccccc}=\int_t^{t+s}\mathbf{E}\left[\left(\mathcal{A}_{\Delta_n} f(u,\rho^{\Delta_n}_u)du-\mathcal{A} f(\rho^{\Delta_n}_u)\right)\prod_{i=1}^m\theta_i(\rho^{\Delta_n}_{t_i})\right]du.\hfill
 \end{multline}
In order to estimate the limit of this term, we have to notice that
\begin{equation}
h(s,\Delta,\rho)=\sqrt{2}\left(h_+(\rho)\cos(\Delta s)+ih_-(\rho)\sin(\Delta s)\right),\end{equation}
for all $\rho\in\mathcal{S}$. It follows that
\begin{multline}
\mathcal{A}_{\Delta_n} f(s,\rho)-\mathcal{A} f(\rho)=\hfill\\
\hphantom{ccccc}=\frac{1}{2}D_\rho^2 f\Big(\sqrt{2}\big(h_+(\rho)\cos(\Delta s)+ih_-(\rho)\sin(\Delta s)\big),\sqrt{2}\big(h_+(\rho)\cos(\Delta s)+ih_-(\rho)\sin(\Delta s)\big)\Big)\hfill\\\hphantom{ccccc=}-\left(\frac{1}{2}D_\rho^2 f(h_+(\rho),h_+(\rho))+\frac{1}{2}D_\rho^2 f(ih_-(\rho),ih_-(\rho))\right)\hfill\\
\hphantom{ccccc}=\frac{1}{2}D_\rho^2 f(h_+(\rho),h_+(\rho))\left(2\cos^2(\Delta s)-1\right)+\frac{1}{2}D_\rho^2 f(ih_-(\rho),ih_-(\rho))\left(2\sin^2(\Delta s)-1\right)\hfill\\\hphantom{ccccc=}
+2D_\rho^2(h_+(\rho),ih_-(\rho))\cos(\Delta s)\sin(\Delta s),\hfill\\
\hphantom{ccccc}=\frac{1}{2}D_\rho^2 f(h_+(\rho),h_+(\rho))\cos(2\Delta s)-\frac{1}{2}D_\rho^2 f(ih_-(\rho),ih_-(\rho))\cos(2\Delta s)\hfill\\\hphantom{ccccc=}
+D_\rho^2(h_+(\rho),ih_-(\rho))\sin(2\Delta s),\hfill
\end{multline}
for all $\rho\in\mathcal{S}$. Then, we get
\begin{eqnarray}
(\ref{35})&\leq&\int_t^{t+s}\mathbf{E}\left[\left(\frac{1}{2}D_\rho^2f(h_+(\rho^{\Delta_n}_u),h_+(\rho^{\Delta_n}_u))\right)\prod_{i=1}^m\theta_i(\rho^{\Delta_n}_{t_i})\right]\cos(2\Delta_n s)du\nonumber\\&&+\int_t^{t+s}\mathbf{E}\left[\left(-\frac{1}{2}D_\rho^2f(ih_-(\rho^{\Delta_n}_u),ih_-(\rho^{\Delta_n}_u))\right)\prod_{i=1}^m\theta_i(\rho^{\Delta_n}_{t_i})\right]\cos(2\Delta_n s)du\nonumber\\
&&+\int_t^{t+s}\mathbf{E}\left[\left(D_\rho^2f(h_+(\rho^{\Delta_n}_u),ih_-(\rho^{\Delta_n}_u))\right)\prod_{i=1}^m\theta_i(\rho^{\Delta_n}_{t_i})\right]\sin(2\Delta_n s)du.
\end{eqnarray}
Let $(i),(ii)$ and $(iii)$ denote the three terms of the sum. We have
\begin{multline}
\vert(i)\vert\leq\hfill\\
\leq\int_{t}^{t+s}\Bigg\vert\mathbf{E}\left[\left(\frac{1}{2}D_\rho^2f(h_+(\rho^{\Delta_n}_u),h_+(\rho^{\Delta_n}_u))\right)\prod_{i=1}^m\theta_i(\rho^{\Delta_n}_{t_i})\right]\hfill\\\hphantom{cccccccccccccccccc}-\mathbf{E}\left[\left(\frac{1}{2}D_\rho^2f(h_+(\mu_u),h_+(\mu_u))\right)\prod_{i=1}^m\theta_i(\mu_{t_i})\right]\Bigg\vert du\hfill\\
\hphantom{c\leq}+\Bigg\vert\textrm{Re}\left(\int_t^{t+s}\mathbf{E}\left[\left(\frac{1}{2}D_\rho^2f(h_+(\mu_u),h_+(\mu_u))\right)\prod_{i=1}^m\theta_i(\mu_{t_i})\right]e^{i2\Delta_n u}du\right)\Bigg\vert.\hfill
\end{multline}
Concerning the first term in this inequality, the term inside the integral $\int_t^{t+s}$ converges to zero, when $n$ goes to infinity (according to the convergence in distribution of $(\rho_t^{\Delta_n})$ to $(\mu_t)$). Moreover, this term is bounded independently of $u$ and $n$ since the processes $(\rho_t^{\Delta_n})$ and $(\mu_t)$ take values in the set of states and the functions are $C_c^2$. By applying the dominated Lebesgue Theorem, the first term converges to zero, when $n$ goes to infinity. Concerning the second term, the term inside the integral in front of the term $e^{i\Delta_n u}$ is bounded and measurable (with respect to $u$), since $\Delta_n$ goes to infinity, a classical result in Fourier transform theory, implies that the integral converges to zero. Let us stress that we can treat the other terms in a similar way and we show that the expression, defined by $(\ref{35})$, converges to zero when $n$ goes to infinity and the result holds.

Now, since the family of processes $(\rho_t^\Delta)$ is tight, we can apply a similar reasoning as the one used to prove the convergence in Theorem $1$. Thus, for all sequences $(\Delta_n)$ converging to infinity, when $n$ goes to infinity, we can extract a subsequence $(\Delta_n')$ such that the property $(\ref{CC1})$ is satisfied for $(\rho_t^{\Delta_n'})$. As a consequence the limit process of $((\rho_t^{\Delta_n'}))$ is a solution for the martingale problem associated with $(\mathcal{A},\rho_0)$. By a similar reasoning as the proof of Theorem $1$, we conclude that the family of processes $(\rho_t^\Delta)$ converges in distribution to $(\rho_t)$ solution of the equation $(\ref{eq2})$.
\end{pf}

\end{document}